\documentclass[a4paper,12pt]{article}
\pdfoutput=1
\usepackage{feynmp-auto,expdlist}
\usepackage{amsmath, amsfonts, amssymb}
\usepackage{graphicx}
\usepackage{enumerate}
\usepackage{hyperref}
\usepackage{latexsym}
\usepackage{dsfont}
\usepackage{hepnicenames}
\usepackage{enumerate}
\usepackage{soul}
\usepackage[normalem]{ulem}
\usepackage{bbold}

\usepackage{units}
\usepackage{mathrsfs,graphicx,rotating,amsmath,amsfonts,mathtools,booktabs,amssymb,wasysym}
\usepackage{hyperref}\usepackage{slashed}
\usepackage[nosort]{cite}
\usepackage[table,xcdraw,dvipsnames]{xcolor}
\usepackage{bm}
\usepackage{graphicx}
\usepackage{multirow,multicol}
\hypersetup{colorlinks,bookmarksopen,bookmarksnumbered,
linkcolor=blus,pdfstartview=FitH,urlcolor=rossos,citecolor=verde}
\allowdisplaybreaks

\renewcommand{\[}{\left[}

\renewcommand{\(}{\left(}
\renewcommand{\)}{\right)}

\def\Lag{\mathscr{L}}

\newcommand{\mio}[1]{}

\def\bpm{\begin{pmatrix}}
\def\epm{\end{pmatrix}}

\usepackage{mathrsfs}

 \newcommand{\fig}[1]{~\ref{fig:#1}}
\newcommand{\sfrac}[2]{#1/#2}

\allowdisplaybreaks
\usepackage{multicol}
\usepackage{color}
\definecolor{rosso}{cmyk}{0,1,1,0.4}
\definecolor{rossos}{cmyk}{0,1,1,0.55}
\definecolor{rossoc}{cmyk}{0,1,1,0.2}
\definecolor{blu}{cmyk}{1,1,0,0.3}
\definecolor{blus}{cmyk}{1,1,0,0.6}
\definecolor{bluc}{cmyk}{1,1,0,0.1}
\definecolor{verde}{cmyk}{0.92,0,0.59,0.25}
\definecolor{verdec}{cmyk}{0.92,0,0.59,0.15}
\definecolor{verdes}{cmyk}{0.92,0,0.59,0.4}

\newcommand{\eq}[1]{~{\rm (\ref{eq:#1})}}

\newcommand{\MeV}{\,{\rm MeV}}

\newcommand{\TeV}{\,{\rm TeV}}

\def\circa#1{\,\raise.3ex\hbox{$#1$\kern-.75em\lower1ex\hbox{$\sim$}}\,}

\newcommand{\beq}{\begin{equation}}
\newcommand{\eeq}{\end{equation}}

\newcommand{\bea}{\begin{eqnarray}}
\newcommand{\eea}{\end{eqnarray}}
\newcommand{\be}{\begin{equation}}
\newcommand{\ee}{\end{equation}}
\font\tenrsfs=rsfs10 at 12pt
\font\sevenrsfs=rsfs7
\font\fiversfs=rsfs5
\newfam\rsfsfam
\textfont\rsfsfam=\tenrsfs
\scriptfont\rsfsfam=\sevenrsfs
\scriptscriptfont\rsfsfam=\fiversfs

\newsavebox\MBox

\renewcommand{\L}{\mathscr{L}}

\def\circa#1{\,\raise.3ex\hbox{$#1$\kern-.75em\lower1ex\hbox{$\sim$}}\,}
\makeatletter

\font\ital=cmu10

\def\hhref#1{\href{http://arxiv.org/abs/#1}{arXiv:#1}}
\usepackage{xstring}
\newcommand{\hhrefq}[1]{\IfSubStr{#1}{:}{\href{http://inspirehep.net/search?ln=en&ln=en&p=#1&of=hb&action_search=Search&sf=&so=d&rm=&rg=25&sc=0}{InSpire:#1}}{\hhref{#1}}}

\def\art{\@ifnextchar[{\eart}{\oart}}
\def\eart[#1]#2#3#4#5#6{{\rm #2}, {\em #3 \bf #4} {\rm (#6) #5} ({\em #1})}
\def\article{\@ifnextchar[{\earticle}{\oarticle}}
\def\oarticle#1#2#3#4#5#6{{\rm #1}, {\ital ``#6''}, {\rm #2 #3 (#5) #4}}
\def\earticle[#1]#2#3#4#5#6#7{{\rm #2}, {\ital ``#7''}, {\rm #3 #4 (#6) #5}  [\hhrefq{#1}]}
\def\hepart[#1]#2{{\rm #2, \sl#1}}
\def\heparticle[#1]#2#3{#2, {\ital ``#3''} [\hhrefq{#1}]}
\newcommand{\doi}[1]{\href{http://dx.doi.org/#1}{[link]}}

\newcommand{\hhrefqq}[1]{\IfBeginWith{#1}{10.}{\href{https://doi.org/#1}{doi:#1}}{\hhrefq{#1}}}
\def\earticle[#1]#2#3#4#5#6#7{{\rm #2}, {\ital ``#7''}, {\rm #3 #4 (#6) #5}  [\hhrefqq{#1}]}
\renewenvironment{thebibliography}[1]
     {\begin{multicols}{2}[\section*{\refname}]%
      \@mkboth{\MakeUppercase\refname}{\MakeUppercase\refname}%
      \list{\@biblabel{\@arabic\c@enumiv}}%
           {\settowidth\labelwidth{\@biblabel{#1}}%
            \leftmargin\labelwidth
            \advance\leftmargin\labelsep
            \@openbib@code
            \usecounter{enumiv}%
            \let\p@enumiv\@empty
            \renewcommand\theenumiv{\@arabic\c@enumiv}}%
      \sloppy
      \clubpenalty4000
      \@clubpenalty \clubpenalty
      \widowpenalty4000%
      \sfcode`\.\@m}
     {\def\@noitemerr
       {\@latex@warning{Empty `thebibliography' environment}}%
      \endlist\end{multicols}}

\newcounter{alphaequation}[equation]
\def\thealphaequation{\theequation\hbox to
0.6em{\hfil\alph{alphaequation}\hfil}}
\def\eqnsystem#1{
\def\@eqnnum{{\rm (\thealphaequation)}}
\def\@@eqncr{\let\@tempa\relax \ifcase\@eqcnt \def\@tempa{& & &} \or
  \def\@tempa{& &}\or \def\@tempa{&}\fi\@tempa
  \if@eqnsw\@eqnnum\refstepcounter{alphaequation}\fi
\global\@eqnswtrue\global\@eqcnt=0\cr}
\refstepcounter{equation} \let\@currentlabel\theequation \def\@tempb{#1}
\ifx\@tempb\empty\else\label{#1}\fi
\refstepcounter{alphaequation}
\let\@currentlabel\thealphaequation
\global\@eqnswtrue\global\@eqcnt=0 \tabskip\@centering\let\\=\@eqncr
$$\halign to \displaywidth\bgroup \@eqnsel\hskip\@centering
$\displaystyle\tabskip\z@{##}$&\global\@eqcnt\@ne
\hskip2\arraycolsep\hfil${##}$\hfil& \global\@eqcnt\tw@\hskip2\arraycolsep
$\displaystyle\tabskip\z@{##}$\hfil
\tabskip\@centering&\llap{##}\tabskip\z@\cr}
\def\endeqnsystem{\@@eqncr\egroup$$\global\@ignoretrue} \makeatother

\definecolor{Gray}{gray}{0.95}

\def\bal#1\eal{\begin{align}#1\end{align}}

\begin{document}
\vspace{1.5cm}

\begin{center}
{\Large\LARGE\Huge \bf \color{rossos}
Cosmological constant:\\
relaxation vs multiverse
%Bottom-less cosmology
}\\[1cm]
{\bf Alessandro Strumia$^{a}$, Daniele Teresi$^{a,b}$}\\[7mm]

{\it $^a$ Dipartimento di Fisica dell'Universit{\`a} di Pisa}\\[1mm]
{\it $^b$ INFN, Sezione di Pisa, Italy}\\[1mm]

\vspace{0.5cm}

{\large\bf\color{blus} 

}

\begin{quote}\large
We consider a scalar field with a bottom-less potential, such as $g^3 \phi$,
finding that cosmologies unavoidably end up with a crunch,
late enough to be compatible with observations if $g \circa{<}1.2 H_0^{2/3} M_{\rm Pl}^{1/3}$.
If rebounces avoid singularities, 
the multiverse acquires new features;
in particular probabilities avoid some of the usual ambiguities.
If rebounces change the vacuum energy by a small enough amount,
this dynamics selects a small vacuum energy and becomes
the most likely source of universes with anthropically small cosmological constant.
Its probability  distribution could avoid the gap by 2 orders of magnitude
that seems left by standard anthropic selection.
\end{quote}

\thispagestyle{empty}
\bigskip

\end{center}

\setcounter{footnote}{0}

\tableofcontents

%Negative $\Lambda$~\cite{0705.0898}.

\section{Introduction}
The vacuum energy $V$ that controls the
cosmological constant 
receives power-divergent
quantum corrections as well as physical corrections of  order
 $M_{\rm max}^4$, where  $M_{\rm max}$ is the mass of the
heaviest particle.
In models with new physics at the Planck scale (e.g.\ string theory) 
one thereby expects Planckian vacuum energies,
and the observed cosmological constant 
(corresponding to the vacuum energy $V_0 \approx (2.3\, {\rm meV})^4$)
can be obtained from a cancellation by one part in $M_{\rm Pl}^4/V_0\sim 10^{120}$.
In tentative models of dimensionless gravity 
the heaviest particle might be the top quark ($M_{\rm max}\sim M_t$,
see e.g.~\cite{1403.4226}), still needing  a
cancellation by one part in $M_{\rm max}^4/V_0 \sim 10^{60}$.

\smallskip

A plausible interpretation of this huge cancellation
is provided by theories with enough vacua such that
at least one vacuum accidentally has the small observed cosmological constant.
Then, assuming that the vacua get populated e.g.\ by eternal inflation,
observers can only develop in those vacua with $V \circa{<}10^3 V_0$~\cite{Weinberg:1987dv}.
More quantitative attempts of understanding anthropic selection 
find that the most likely vacuum energy measured by a random observer
is about 100 times larger
that the vacuum energy $V_0$ we observe~\cite{Weinberg:1987dv,astro-ph/9908115,astro-ph/0005265,0705.0898}
(unless some special measure is adopted, for instance as in~\cite{0805.2173,hep-th/0605263,hep-th/0702115,0808.3770}).
This mild remaining discrepancy might signal some missing piece of the puzzle. 

%Theories with $N$ scalars can easily have $2^N$ vacua.

Recently \cite{Graham:2019bfu} (see also~\cite{1608.05715})
proposed a cosmological model that could make the
cosmological constant partially smaller and negative.  It needs two main ingredients:
\begin{enumerate}
\item[a)] `Rolling': a scalar field $\phi$ with a quasi-flat potential and no bottom
(at least in the field space probed cosmologically), such as
$V_\phi =- g^3\phi$ with small $g\circa{<}H_0^{2/3} M_{\rm Pl}^{1/3}$
where $H_0$ is the present Hubble constant.
\listpart{Then, a cosmological phase during which the energy density is
dominated by $V_\phi$ 
(with a value such that 
$\phi$ classically rolls down its potential)
ends up when $V_\phi$ crosses zero and becomes
slightly negative, starting contraction. During the contraction phase the kinetic energy of $\phi$ rapidly blue-shifts and, assuming some interaction with extra states, gets converted into a radiation bath, thus reheating the Universe and maybe triggering the following dynamics.}

%\item[b)] a thermal bath that can store the $\phi$ kinetic energy slowering its rolling;
\item[b)] `Rebouncing': a  mechanism that turns a contracting universe into an expanding universe. Furthermore, to get a small positive (rather than negative) cosmological constant,
the authors of
 \cite{Graham:2019bfu} assume multiple minima and a `hiccupping' mechanism that populates
vacua up to some energy density $V_{\rm rebounce}$.
\end{enumerate}
%The mechanism works as follows: 
%By assumption b) a rebounce produces a universe with a slightly larger, positive, cosmological constant.
Hence, at this stage the Universe appears as hot, expanding and with a small positive cosmological constant, i.e.~with standard hot Big-Bang cosmology.
In this way, the cancellation needed to get the observed cosmological constant gets partially
reduced by some tens of orders of magnitude, such that theories
with $M_{\rm max}\sim \MeV$ no longer need accidental cancellations\footnote{Notice that the mechanism of~\cite{1608.05715} can relax vacuum energies up to $\sim \TeV^4$, while having a cutoff in the MeV range or lower.}~\cite{1608.05715,Graham:2019bfu}. 
However particles almost $10^6$ heavier than the electron exist in nature.

The authors of~\cite{Graham:2019bfu} restricted the parameter space
of their model in order to avoid eternal inflation.
However other features of the Standard Model, in particular light fermion masses,
suggest that anthropic selection is playing a role~\cite{mq}.
The weak scale too might be anthropically constrained~\cite{1906.00986}.
Taking the point of view that a multiverse remains needed,
we explore  the role that the above ingredients a) and b),
assumed to be generic enough,
might play in a
multiverse context.
Is an anthropically acceptable  vacuum more easily found by random chance or through the
mechanism of~\cite{Graham:2019bfu}?

In section~\ref{phi} we consider in isolation the ingredient a),
finding that all observers
eventually end up in an anti-de-Sitter crunch,
that can be late enough to be compatible with cosmological data.
%an observer in a universe dominated by matter, radiation or 
%positive vacuum energy .
In section~\ref{tmul} we consider in isolation the ingredient b),
finding that it modifies the multiverse structure, in particular leading
to multiple cycles of a ``temporal multiverse''.

Adding both ingredients a) and b), in section~\ref{RR}
we show that the mechanism of~\cite{Graham:2019bfu}
can have a dominant multiverse probability of forming
universes with an anthropically acceptable vacuum energy.
In such a case, the small discrepancy
left by usual anthropic selection (the measured vacuum
energy $V_0$ is 100 times below its most likely value)
can be alleviated or avoided.
Conclusions are given in section~\ref{concl}.

%Its relevance for the anthropic selection of the cosmological constant
%depends on how strong is the hiccupping ingredient.

\section{Rolling: a bottom-less scalar in cosmology}\label{phi}
A scalar potential with a small slope but no bottom is one of the ingredients
of~\cite{Graham:2019bfu}.
We here study its cosmology irrespectively of the other ingredients.
We consider a scalar field $\phi$ with  Lagrangian 
\begin{equation}
\L_\phi = \frac{(\partial_\mu \phi)^2 }{2}- V_\phi(\phi) ,
\end{equation}
where the quasi-flat potential can be approximated as $V_\phi(\phi)\simeq  -g^3 \phi$ with small $g$.
We consider a flat homogeneous universe with scale-factor $a(t)$
(with present value $a_0$)
in the presence of $\phi$ and of non-relativistic matter with density $\rho_m(a)=\rho_m(a_0) \sfrac{a_0^3}{a^3}$,
as in our universe at late times.
Its cosmological evolution is described by the following equations
\begin{eqnsystem}{sys:uni}
\frac{\ddot a}{a} &=& -\frac{4\pi G}{3} (\rho + 3p)\\
\ddot \phi  &=& - 3 \frac{\dot a}{a} \dot \phi -V_\phi'
\end{eqnsystem}
where $G=1/M_{\rm Pl}^2$ is the Newton constant;
$\rho =\rho_\phi+\rho_m$ and $p=p_\phi$ are
the total energy density and pressure with
\beq
\rho_\phi= \frac{\dot\phi^2}{2}+V_\phi ,\qquad
 p_\phi = \frac{\dot\phi^2}{2}-V_\phi .
\eeq
In an inflationary phase with negligible radiation and matter density $\rho_m$
the scale factor grows as $a\propto e^N$
and $\phi$ undergoes classical slow-roll 
$\dot\phi \simeq - V'_\phi/3H$ i.e.\
$d\phi/dN \simeq -V'_\phi/3H^2$ 
as well as quantum fluctuations $\delta \phi\sim H/2\pi$ per $e$-fold,
where $H^2 = {8\pi} V/3M_{\rm Pl}^2$.
We assume that all other scalars eventually settle to their minimum,
such that we can assume $V = V_\phi$, up to a constant that can be reabsorbed in a shift in $\phi$. 

Classical motion of $\phi$ dominates over its quantum fluctuations
for field values such that $|V'_\phi| \gg H^3$.
The critical point is $ \phi_{\rm class} \sim -M_{\rm Pl}^2/g$
which corresponds to vacuum energy $V_{\rm class} \sim g^2 M_{\rm Pl}^2$.
Classical slow-roll ends when $V_\phi \sim \dot\phi^2$:
this happens at $\phi\sim \phi_{\rm end}\sim M_{\rm Pl}$
which corresponds to $V_\phi \sim V_{\rm end}\sim- g^3 M_{\rm Pl}$.
Such a small $V_\phi \approx 0$ is a special point of the cosmological
evolution when $V_\phi$ dominates the energy density~\cite{1608.05715,Graham:2019bfu}.
The scale factor of an universe dominated by $V_\phi$ expands by
$N \sim M_{\rm Pl}^2/g^2$ $e$-folds
while transiting the classical slow-roll region.

\smallskip

Eternal inflation occurs for field values such that $V_\phi \circa{>}V_{\rm class}$: 
starting from any given point $\phi < \phi_{\rm class}$ 
the field eventually
fluctuates down to $ \phi_{\rm class}$
after $N \sim| \phi| M_{\rm Pl}^2/g^3 $ $e$-folds.
The Fokker-Planck equation for the probability  density $P(\phi,N)$ in comoving coordinates
of finding the scalar field at the value $\phi$ has the form of a leaky box~\cite{astro-ph/9401042} 
\beq
 \displaystyle \frac{\partial P}{\partial t} =  \frac{\partial }{\partial \phi}\bigg( 
 \frac{M_{\rm Pl}^2}{4\pi}  \frac{\partial H}{\partial \phi} P +
 \frac{H^{3/2}}{8\pi^2} \frac{\partial }{\partial \phi}(H^{3/2} P)\bigg) .
 \label{Fokker}
 \eeq
%\beq
% \displaystyle \frac{\partial P}{\partial t} = \frac{H^3}{8\pi^2}  \frac{\partial^2 P }{\partial \phi^2} +
% \frac{\partial }{\partial \phi}  \bigg(\frac{V'}{3H}   P\bigg) ,
% \label{Fokker}
% \eeq
This equation admits stationary solutions
where $P$ decreases going deeper into the quantum region
(while being non-normalizable),
and leaks into the classical region. 
%where the first term is a constant 
%(solution up to the derivative in $H$)
% with a peak in the classical region. 
% The other solution grows at $\phi\to\infty$.

\smallskip

\begin{figure}[t]
$$\includegraphics[width=\textwidth]{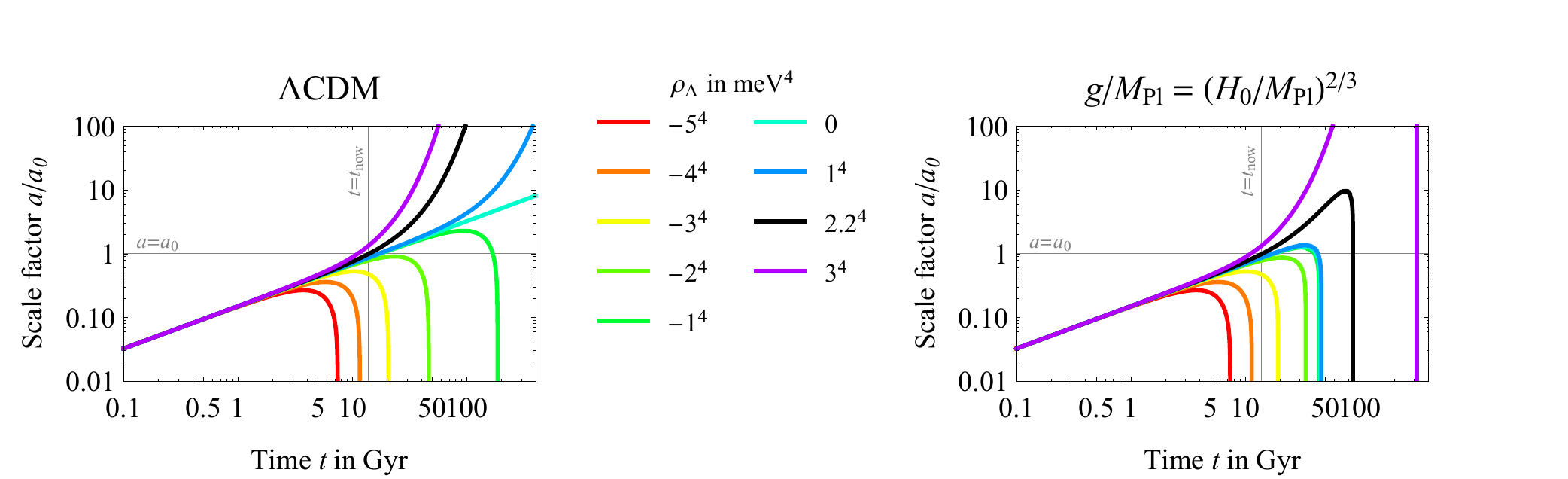}$$
\caption{\em 
\label{fig:MatCosmo} 
We consider a flat universe with matter fixed to its observed density.
{\bf Left:} evolution of the scale factor (inverse of the temperature)
for different cosmological constants.
\label{fig:nofriction} 
{\bf Right:} evolution of the scale factor 
in the presence of a scalar $\phi$ with bottom-less potential $g\phi^3$, initially fixed at different cosmological constants.
}
\end{figure}

%\begin{figure}[t]
%$$\includegraphics[width=0.65\textwidth]{figs/nofriction}$$
%\caption{\em 
%}
%}
%\end{figure}

\begin{figure}[t]
$$\hspace{-2em}\includegraphics[width=0.65\textwidth]{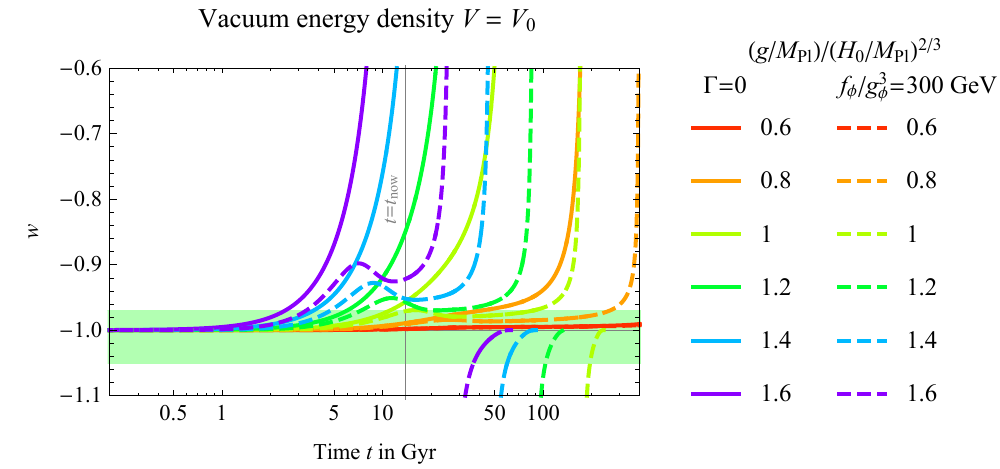} \quad \includegraphics[width=0.27\textwidth]{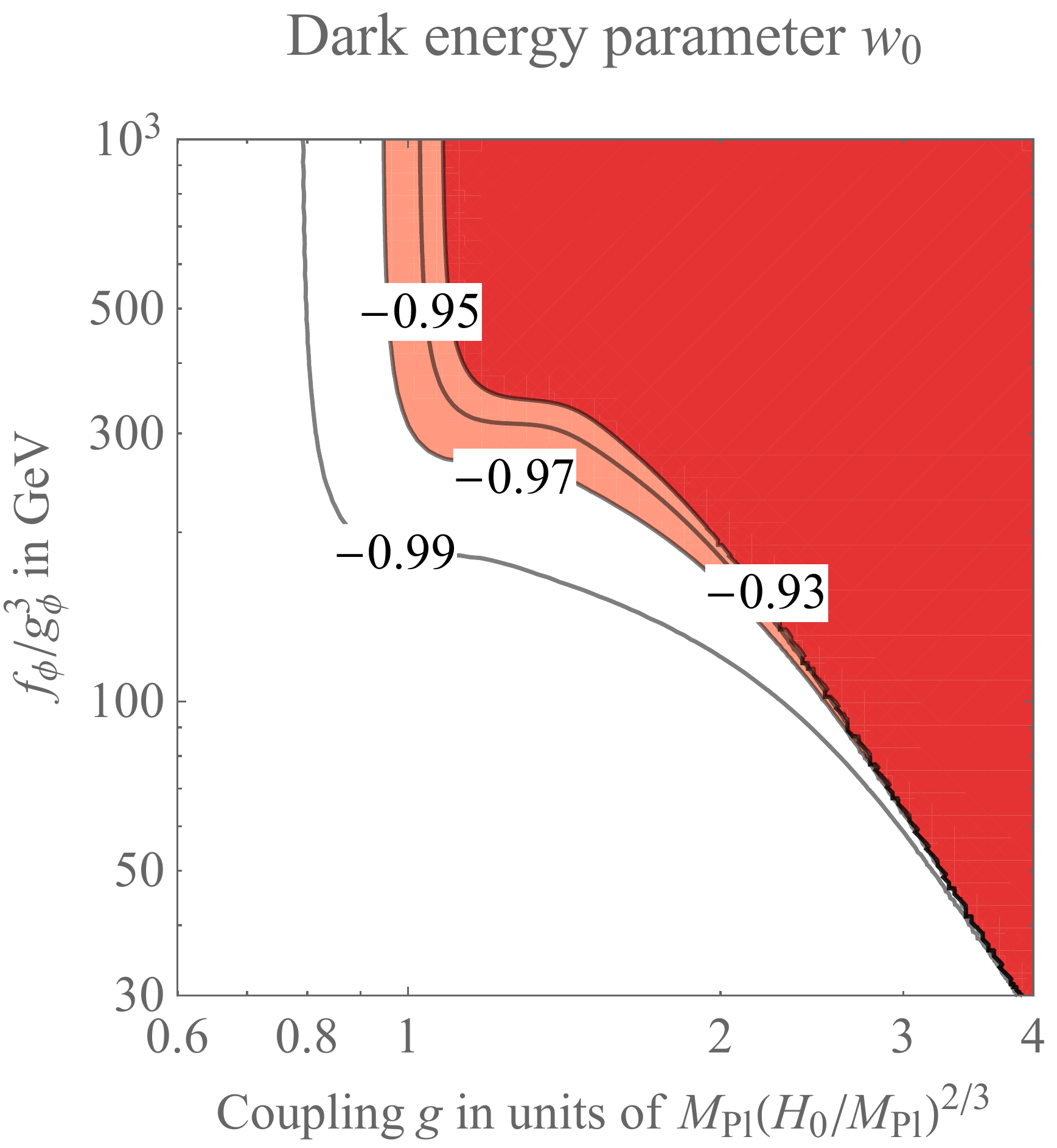}$$
\caption{\em 
\label{fig:w} {\bf Left}:
We consider cosmologies that reproduce, at early times,
the measured vacuum energy density $V_0$,
for different values of the slope parameter $g$.
We plot the time evolution of the dark-energy parameter $w=p_\phi/\rho_\phi$. We consider both models with thermal friction (dashed curves) and without thermal
friction (continuous curves, $\Gamma=0$). 
{\bf Right}: iso-contours of $w$ today.
The shaded regions are disfavoured at 1 and 2 standard deviations by current data.}
\end{figure}

A large density $\rho$ of radiation and/or matter is present during the early big-bang phase.
The scalar $\phi$, similarly to a cosmological constant, is irrelevant during this phase.
The variation in the scalar potential energy due to its slow-roll is negligible as long as
\beq\label{eq:gbound}
 |V'_\phi| \ll H^2 M_{\rm Pl}.\eeq 
Indeed
\beq \frac{d V_\phi}{d N} = V'_\phi \frac{d\phi}{dN} = \frac{V_\phi^{\prime 2}}{3H^2} \ll \rho \sim H^2 M_{\rm Pl}^2.\eeq
Thereby the evolution of a scalar field with a very small slope $g^3$ 
becomes relevant only at late times %when $1/H$ becomes very large, or equivalently
when the energy density $\rho$ becomes small enough,
$\rho\circa{<} V_\phi$.
%$g^3/M_{\rm Pl}^3 \sim H^2/M_{\rm Pl}^2 \sim \rho/M_{\rm Pl}^4$.

\medskip

Fig.\fig{nofriction} shows the cosmological evolution of our universe,
assuming different initial values
of the vacuum energy density $V_\phi(\phi_{\rm in})$.
%with a negative vacuum energy $\rho_\Lambda = -\Lambda^4$
%and matter (as in our universe at late times),
%such that the total energy density is
%$\rho=\rho_\Lambda (1-a^3_{\rm max}/a^3)$
%and $H^2 = {8\pi G}\rho/3$.
%
If such vacuum energy is negative, a crunch happens roughly as in standard cosmology, after a time
\begin{equation}\label{eq:crunch_time}
t_{\rm crunch} = 2 \int_0^{a_{\rm max}} \frac{da}{a H}=\sqrt{\frac{\pi}{6}}
\frac{M_{\rm Pl}}{\sqrt{-V_\phi(\phi_{\rm in})}}\approx  \sqrt{\frac{V_{0}}{-V_\phi(\phi_{\rm in})}}
%\(\frac{\Lambda_{\rm our}}{\Lambda}\)^2 
\, 3.6 \times \unit[10^{10}]{yr}.
\end{equation}
Unlike in standard cosmology
the Universe finally undergoes a crunch even if $V_\phi(\phi_{\rm in})\ge 0$, because $\phi$ starts 
dominating the energy density (like a cosmological constant) and rolls down
(unlike a cosmological constant).
The crunch happens in the future
for the observed value of the cosmological constant and for the
value of $g^3 = H_0^{2} M_{\rm Pl}$ assumed in fig.\fig{nofriction}.
Here $H_0$ is the present Hubble constant.
For larger $g$ the crunch happens earlier.
We do not need to show plots with different values of $g$ because,
up to a rescaling of the time-scale, the cosmological evolution only depends on 
$g^3/H^2_* M_{\rm Pl}$ where $H_*$ is the Hubble
constant when matter stops dominating
(in our universe, $H_*$ is the present Hubble constant $H_0$ up to order one factors). 
This means that large enough values of the vacuum energy
behave as a cosmological constant for a while.

Fig.\fig{w} shows the time evolution of the dark-energy parameter $w=p_\phi/\rho_\phi$
for cosmologies that reproduce
the present value of the matter and dark energy densities 
and for different values of the slope parameter $g$.
The observed present value $w_0= -1.01 \pm 0.04$~\cite{1807.06209,PDG} implies the
experimental bound
$g \circa{<}1.2 H_0^{2/3} M_{\rm Pl}^{1/3}$. 
This means that
the anthropic restriction on the vacuum energy remains essentially
the same as in standard cosmology (where vacuum energy is a cosmological constant),
despite that all cosmologies (even for large positive cosmological constant)
eventually end with a crunch.

%. \DT{For $\Lambda$ positive and $g$ sufficiently large the finite life-time of the universe modifies anthropic considerations, in particular the number of observers created in an Universe with $\Lambda$. In the extreme case that the lifetime is less than $\approx 8.5$ billion years no observers can be formed. I will take this into account, approximately in the analysis of the time multiverse.}

\subsection{Cosmology with a falling scalar and thermal friction}
The mechanism of~\cite{Graham:2019bfu} employs some interaction
that, during the crunch, converts the kinetic energy $\dot\phi^2/2$
into a thermal bath by particle production due to the evolution of $\phi$, thus reheating the Universe up to potentially high temperatures.
The scalar $\phi$ can have interactions compatible with its lightness.
Indeed, $\phi$ might be a Goldstone boson with derivative interactions
e.g.\ to extra vectors $F_{\mu\nu}$ or fermions $\Psi$
\beq\Lag_{\rm bath} =\frac{\phi}{f_\phi} F_{\mu\nu}\tilde F_{\mu\nu} + y \phi \bar \Psi\gamma_5 \Psi.\eeq
In the presence of a cosmological thermal bath of these particles $\phi$ acquires a thermal friction
$\Gamma$
without acquiring a thermal mass~\cite{Graham:2019bfu}.
For example $\Gamma\sim 4g_\phi^6 T^3_{\rm bath}/\pi f_\phi^2$ in the bath
of vectors with gauge coupling $g_\phi$.
We study  its effects during the expansion phase.
The cosmological equations of eq.~(\ref{sys:uni}) generalise to
\begin{eqnsystem}{sys:uni}
\frac{\ddot a}{a} &=& -\frac{4\pi G}{3} (\rho + 3p)\\
\ddot \phi  &=& - (3 \frac{\dot a}{a}+\Gamma) \dot \phi -V_\phi' \label{eq:EoM_Gamma}\\
\dot \rho_{\rm bath} &=& -3\frac{\dot a}{a} (\rho_{\rm bath}+p_{\rm bath}) + \Gamma \dot\phi^2 \label{eq:drhoRH}
\end{eqnsystem}
where $\rho$ and $p$ are
the total energy density and pressure
\beq
\rho = \rho_\phi+\rho_m+\rho_{\rm bath},\qquad
p =p_\phi + p_{\rm bath}.
\eeq
 Equation\eq{drhoRH}, dictated by energy conservation,
tells the evolution of the energy density of the bath $\rho_{\rm bath}$
in view of the expansion of the universe and of the energy injection from $\phi$.
The  pressure $p_{\rm bath}$
 equals $\rho_{\rm bath}/3$ (0) for a relativistic (non-relativistic) bath.

\smallskip

The presence of a bath can modify the expansion phase,
even adding a qualitatively new intermediate period during which
$\phi$ rolls down the potential acquiring
an asymptotic  velocity $\dot \phi \sim V'_\phi/\Gamma(T_{\rm bath})$ 
while the bath, populated by the $\phi$ kinetic energy,
acquires a corresponding quasi-stationary temperature $T_{\rm bath} \sim
(g^6/H\Gamma)^{1/4}$.
The final crunch gets delayed but it eventually happens
as illustrated by the dashed curves in fig.\fig{w}, and as we now
show 
analytically.
Conservation of `energy' gives a
 first integral of eq.s~(\ref{sys:uni}) well known as
%if the collapse generically occurs also in its presence. 
% The latter acts against the collapse of the Universe. Nevertheless, we show in Fig.~\ref{fig??} \DT{what should I plot precisely?} that the  generically occurs also in the presence of a sizeable $\Gamma$, while a thermal bath is indeed produced, for instance of vectors.
%This can also be understood  as follows. 
Friedmann's equation, $
H^2 = \sfrac{8 \pi G \rho}{3} $.
By differentiating it and using~\eqref{eq:EoM_Gamma} and \eqref{eq:drhoRH} one obtains
%\be 
%\dot H = - 4 \pi G \( \dot \phi^2 + \rho_{\rm bath} + p_{\rm bath}\) = - 4 \pi G \(\rho_{\phi} + \rho_{\rm bath} + p_{\phi} + p_{\rm bath} \) \leq 0
%\ee
\be 
\dot H = - 4 \pi G \( \rho+p \) \leq 0
\ee
in which the contribution of $\Gamma$ cancels. 
In general, $\dot H$ is non-positive because the null-energy condition $\rho + p \geq 0$
is satisfied. 
At the turning point $H=0$ one has $\rho = 0$
thanks to a cancellation between a positive $\rho_{\rm bath}$
and negative $\rho_\phi$:
for our system this implies $p>0$ such that
$\dot H$ %would require $p = p_\phi + p_{\rm bath} = 0$ which is not possible for $\dot \phi^2 >0$ and $w > -1$. Therefore $\dot H$
 is strictly negative and the Universe starts collapsing.
 This avoids the Boltzmann-brain paradox that affects
{cosmologies with positive cosmological constant}~\cite{hep-th/0208013,hep-th/0510003,hep-th/0610132}. 
The right panel of fig.\fig{w} shows that
 interactions relax the observational bound on $g$ by an amount proportional
 to $f_\phi^{-1/3}$, for $g$ large enough.
    
 %For all values of $\Lambda$ the Universe will eventually recollapse. 
% \xxx{AS: even if the bath is matter?} \DT{S\`i, se risolvo il sistema $H=0$, $\dot H =0$ mi viene $w = - (\dot\phi^2+\rho_{\rm bath})/\rho_{\rm bath}$ strettamente minore di -1 se $\dot \phi \neq 0$. L'unico loophole \`e la possibili\`a $\dot \phi =0$ che darebbe a quel punto un universo che raggiunge uno stato statico. Nel nostro caso questo non avviene perch\`e $\dot\phi$ ha una velocit\`a terminale data da Hubble o $\Gamma$ friction.}
 
%\xxx{too strong?} \DT{Se vince non soffre di ambiguit\`a; all'interno di un singolo universo c'\`e un concetto di tempo ben definito. L'unica regolarizzazione (misura) ragionevole \`e: consideriamo l'evoluzione fino a un tempo $t$; calcoliamo la probabilit\`a (eventi finiti); mandiamo il cutoff $t \to \infty$. Non vedo altre possibilit\`a non artificiose.}
 
%\beq\label{eq:gbound}
% |V'_\phi| \ll H^2 M_{\rm Pl} \sqrt{1+\Gamma/3H}.\eeq
%This allows to increase the slope $g^3$, see eq.\eq{gbound}.
%
%Notice that $\dot H =- G (\rho+p)$ [O(1) omitted]: as before $p\neq0$ at $\rho=0$
%allows to cross $H=0$.

%\AS{AS: off-topic:
%it would be interesting to see if one can get from $m_e$ to $m_t$.
%And to study if the ghost can rebounce. And super-Planckian...
%}

% hep-th/0204168,
\section{Rebouncing: a temporal multiverse}\label{tmul}
It is usually assumed that anti-de Sitter
regions with negative vacuum density collapse to a big-crunch singularity.
The resolution of this singularity is not known (for example in perturbative string theory~\cite{hep-th/0206228}),
so it makes sense to consider the opposite possibility b):
that collapsing anti-de-Sitter regions rebounce into an expanding space.
The mechanism of~\cite{Graham:2019bfu} assumes that
the vacuum energy density changes by a small amount $V_{\rm rebounce}$
in the process.

Following the usual assumptions that anti-de Sitter vacua are `terminal',
various authors tried to compute the statistical distribution of vacua in a multiverse
populated by eternal inflation,
in terms of vacuum decay rates $\kappa_{IJ}$ from vacuum $I$ to  vacuum $J$~\cite{Garriga:2005av}.
These rates are defined up to unknown multiverse factors, because
eternal inflation gives an infinite multiverse, so
probabilities are affected by divergences.
Some measures lead to paradoxes (see for instance~\cite{hep-th/0208013,hep-th/0510003,hep-th/0610132,hep-th/0702178}).
Furthermore, even if the multiverse statistics were known, 
its use would be limited by our ability of observing only one event (our universe).
Despite these drawbacks and difficulties
many authors tried addressing the issue (see e.g.~\cite{Garriga:2005av,0808.3778,0812.0005,0901.4806,1005.2783,1104.2324}).

%The above attempts assume that anti-de Sitter vacua are `terminal':
%observers who get there collapse to a big-crunch singularity.

If vacua with negative cosmological constant are not terminal,
multiverse dynamics would change as follows (see also~\cite{0901.2644,1210.7540}).
For simplicity we consider a toy multiverse with 3 vacua: 
$S$ (de Sitter), $M$ (Minkowski) and  $A$ (anti de Sitter).
The evolution of the fraction of `time' spent by an `observer' in the vacua is described by
an equation of the form (see e.g.~\cite{Garriga:2005av})
\beq
\frac{d}{dt} \begin{pmatrix}
f_S \cr f_M \cr f_A
\end{pmatrix}
=
\left[
\begin{pmatrix}
-\kappa_{SM} -\kappa_{SA} & 0 & 0\cr
\kappa_{SM} & - \kappa_{MA} &0\cr
\kappa_{SA} & \kappa_{MA} & 0
\end{pmatrix}+
\begin{pmatrix}
0& 0 & \kappa_{AS}\cr
0& 0 & \kappa_{AM}\cr
0&0&-\kappa_{AS}-\kappa_{AM}
\end{pmatrix}\right]
\cdot\begin{pmatrix}
f_S \cr f_M \cr f_A
\end{pmatrix}
\eeq
%where the $\kappa_{ij}$ are related to the decay rates from $i$ to $j$,
If anti-de-Sitter vacua are terminal only the first term is present:
then, in a generic context, the frequencies $f_I$ are dominated 
by decays from the most long-lived de Sitter vacuum~\cite{Garriga:2005av}.
If anti-de-Sitter vacua are not terminal, 
the second term containing the $\kappa_{AJ}$ recycling
coefficients is present, allowing for 
a steady state solution 
\beq
\begin{pmatrix}
f_S \cr f_M \cr f_A
\end{pmatrix}\propto
\begin{pmatrix}
\kappa_{AS} \kappa_{MA} \cr
\kappa_{AM} \kappa_S+ \kappa_{AS}\kappa_{SM}\cr
\kappa_{MA} \kappa_S
\end{pmatrix}
\eeq
where $\kappa_S =\kappa_{SA}+\kappa_{SM}$. 

If anti-de-Sitter crunches rebounce due to some generic mechanism when
they reach Planckian-like energies,
the $\kappa_{AJ}$ coefficients might be universal and populate all lower-energy vacua.
The mechanism of~\cite{Graham:2019bfu} needs a milder hiccupping mechanism,
that only populates vacua with vacuum energy slightly higher 
than the specific AdS vacuum that crunched.

\section{Rolling and rebouncing: the hiccupping multiverse}\label{RR}
Finally we consider the combined action of the ingredients a) and b) of~\cite{Graham:2019bfu}.
%a scalar with a bottom-less potential together with collapsing universes  rebouncing into expanding
%universes with vacuum energy varied by $\sim V_{\rm rebounce}$.

Due to a), observers in de Sitter regions unavoidably end up sliding down the $V_\phi$ potential 
until the vacuum energy becomes small and negative, of order $V_{\rm end}\sim - g^3 M_{\rm Pl}$.
This happens even if the vacuum energy is so big that quantum fluctuations of $\phi$ initially
dominate over its classical slow-roll.
This process can end de Sitter faster than quantum tunnelling to vacua with lower energy densities,
as the vacuum decay rates are exponentially suppressed by possibly large factors.
%\DT{Assuming that bounce and hiccuping are described in previous sections, otherwise add here}
%Let us consider the process of universe formation in the multiverse. 
%We assume that the standard spatial multiverse is generated by eternal inflation \DT{generated by $V_\phi$?}. At some point a quantum fluctuation creates a patch where $\phi$ is small enough that classical motion dominates; in this pocket universe $\phi$ rolls down and, because of the mechanism of~\cite{Graham:2019bfu}, the cosmological constant gets relaxed to a value $\Lambda_{\rm fin} \sim - g^3 M_{\rm Pl}$. 

Due to b), contracting regions with small negative vacuum energy density $\sim V_{\rm end}$
eventually rebounce, becoming expanding universes with
vacuum energy varied by $\sim V_{\rm rebounce}$.
As in the previous section, a \emph{temporal} multiverse is created: different values of the vacuum energy are sampled in different cycles. 

A new feature arises due to the presence of both ingredients: 
all cycles now last a finite time.
A single patch samples different values of physical parameters (for example vacuum energies) with a given probability distribution.
We refer to this as temporal multiverse.

%\DT{This is an important novel feature that allow to discuss probabilities on firm grounds; O comunque qualche frase che stressa, rischiamo un under-claim; a me sembra una cosa molto importante.}

%\footnote{There is a preferred way of enumerating the different instances of $\Lambda$, by simply following the evolution in time. Of course one could artificially change the way the different instances of $\Lambda$ are enumerated and obtain an ambiguous probability, but this looks extremely byzantine \DT{find possible better word} in the current setting.}.

\smallskip

A {\em disordered} temporal multiverse arises if the extra vacuum energy generated during the rebounce, $V_{\rm rebounce}$,
is typically larger than $V_{\rm class}$,
such that the small vacuum energy selected by the rolling mechanism at the previous cycle is lost.
Small vacuum energy is not a special point of this dynamics, and the usual anthropic selection argument 
discussed by Weinberg~\cite{Weinberg:1987dv}  applies:
the most likely value of the cosmological constant is 2 orders of magnitude
above its observed value
(for the observed value of the amount of primordial inhomogeneities, $\delta \rho/\rho\sim10^{-5}$).
This discrepancy by 2 orders of magnitude (or worse if $\delta \rho/\rho$ can vary)
possibly signals that anthropic selection is not enough to
fully explain the observed small value of the cosmological constant.
%Usually, this is instead achieved by choosing the geometry of the measure such that for relatively large values of $\Lambda$ only observers in a smaller volume are counted \DT{read better causal patch/diamond measures.}.

An \emph{ordered} temporal multiverse arises if, instead, 
the contraction/bounce phase changes  the vacuum energy density by an amount $V_{\rm rebounce}$
smaller 
enough than $V_{\rm class}$ such that,
when one cycle starts, it proceeds forever giving rise at some point
to an anthropically acceptable vacuum energy.
We refer to this possibility as `hiccupping'.
A small vacuum energy $\sim V_{\rm end}$
is a special point of this dynamics, such that, depending on details of hiccupping,
the probability distribution of the vacuum energy can peak below the maximal value allowed by anthropic selection. In the absence of knowledge of a clear rebouncing mechanism (while some recent attempts have been done, see e.g.~\cite{Graham:2017hfr}), having one rather than the other possibility shoud be considered as an assumption.

Going beyond the two limiting cases discussed above,
an intermediate situation can be broadly characterised by different scales:
$V_{\rm rebounce}$ (the amount of randomness in vacuum energy at each rebounce);
$V_{\rm class}\sim g^2 M_{\rm Pl}^2$ (the critical value of $V_\phi$
above which $\phi$ starts fluctuating);
$V_{\rm max} \sim M_{\rm max}^4$ (the maximal energy scale in the theory).
The observed  value $V_0$ of the vacuum energy
can either be reached trough the rolling dynamics of~\cite{Graham:2019bfu}
or by the usual random sampling the multiverse.
The relative probability of these two histories is
\beq \frac{\wp_{\rm rolling}}{\wp_{\rm random}}
\sim \frac{\min [1, (V_{\rm class}/V_{\rm rebounce})^{V_{\rm rebounce}/V_0}] }{V_0/V_{\rm max}}.\eeq
We ignored possible fine structures within each scale.
If the mechanism of~\cite{Graham:2019bfu}
provides the dominant source of anthropically-acceptable vacua
(those with $ V \circa{<} 10^3 V_0$),
the observed value $V_0$ of the vacuum energy density can have a probability
larger than in the usual multiverse scenario.

\medskip

\subsection{Possible hiccuping dynamics}
Let us now discuss the value of  $V_{\rm rebounce}$ from a theoretical point of view.
If the rebounce occurs when the contracting region heats up to temperatures $T_{\rm rebounce}$
(or, more in general, energy density $\sim T_{\rm rebounce}^4$),
one can expect that scalars lighter than $T_{\rm rebounce}$ 
can jump to different minima
(assuming that  potential barriers are characterised by the mass).
If the rebounce happens when contraction reaches Planckian densities, one expects $V_{\rm rebounce}\sim M_{\rm Pl}^4$.
A very small value of $V_{\rm rebounce} \sim V_0$ could arise assuming that
%Let us now consider hiccupping more in detail. At each cycle, relaxation occurs and eventually brings the cosmological constant down to the value $V_{\rm end} \approx - g^3 M_{\rm Pl}$. Thus, if $g/M_{\rm Pl} \ll (H_0/M_{\rm Pl})^{2/3}$, $V \simeq 0$ is a special point of the dynamics. Then, the rebounce occurs at some temperature $T_{\rm bounce}$ and the hiccupping changes the cosmological constant to a new value $V_{\rm end} + \delta V \simeq \delta V$. 
the contraction/rebounce/expansion phase triggers movement 
of some lighter fields $\phi'$ with potentials such that
vacua close by in field space have similar energies.
`Ordered' landscapes of this kind have been considered, for instance, in~\cite{Abbott:1984qf,1504.07551,1609.06320,1801.03926,1809.07338,1811.12390,1904.00020}.\footnote{In our context,
this property must only be obeyed by some lighter fields, not by the full landscape.}
An example of this hiccupping structure is provided by Abbott's model~\cite{Abbott:1984qf}, i.e.~a 
light scalar $\phi'$ with potential that can be (at least locally) approximated as
\be 
V_{\phi'} = -g_{\phi'}^3 \phi' - \Lambda^4 \cos \frac{\phi'}{f_{\phi'}} 
\ee
with, again, a very small slope $g_{\phi'}^3$, such that $g_{\phi'}^3 f_{\phi'} \ll V_0$.\footnote{Such small slopes are typical, for instance, of relaxion models~\cite{1504.07551}, where they can be generated dynamically by the clockwork mechanism, 
% bleeeeeeah
either in its multi-field~\cite{1511.00132,1511.01827} or extra-dimensional~\cite{1610.07962,1802.01591} versions.}
During a given cycle the field $\phi'$ is trapped in a local minimum  (that we may take at $\phi'=0$) provided that $\Lambda$ is large enough to quench tunnelling. 
At the end of the cycle, during the contraction/rebounce/expansion phase, the barriers become irrelevant for some time and the field $\phi'$
is free to diffuse from $\phi' = 0$ by thermal or de Sitter fluctuations.
We focus on de Sitter fluctuations, given that
a phase of the usual inflation with Hubble constant $H_{\rm infl}$
is probably needed to explain the observed primordial inhomogeneities.
For $g_{\phi'} \ll H_{\rm infl}$ the quantum evolution dominates 
(the classical rolling of $\phi'$ gives a negligible variation in $ V$, of order $\sim N_{\rm infl} g_{\phi'}^6/H_{\rm infl}^2$)
and the field $\phi'$ acquires a probability density 
$\mathcal{P}_1 \sim \exp(- 2 \pi^2 \phi^{\prime2}/H^2_{\rm infl}  N_{\rm infl})$, where
$N_{\rm infl}$ is the number of $e$-folds of inflation. 
Hence, when barriers become relevant again, the vacuum energy has probability density
\be \label{eq:ordhic}
\mathcal{P}_1(V) \sim \exp\(-  \frac12 \frac{V^2}{V_{\rm rebounce}^2}\) \qquad\hbox{with}\qquad
V_{\rm rebounce}=g_{\phi'}^3  \frac{H_{\rm infl}}{2 \pi} \sqrt{N_{\rm infl}}.
\ee
Quantum fluctuations happen differently in
different Hubble patches: 
after inflation regions in different vacua progressively return in causal contact,
and the region with lower vacuum energy density
expands into the other regions.
If $f_{\phi'} \ll H_{\rm infl} $ there is a order unity probability that this is
happening now on horizon scales,
giving rise to gravitational waves~\cite{1703.02576}
(and to other signals as in~\cite{1205.6260} if $\phi$ couples to photons).
The field $\phi'$ (for $\Lambda=0$)
can be identified with $\phi$ provided
that $N_{\rm infl} \circa{>} \(\sfrac{2 \pi M_{\rm Pl}}{H_{\rm infl}}\)^2 $
is large enough that $V_{\rm rebounce} \geq |V_{\rm end}|$. 
The above hiccup mechanism can be part of the
scenario of \cite{Graham:2019bfu}, that tries avoiding the multiverse.

%\DT{Notice that there is the minimal possibility to identify $\chi$ with $\phi$: the same field  relaxes the cosmological constant and is responsible for the hiccupping. Given that to obtain a positive vacuum energy one needs $V_{\rm rebounce} \geq |V_{\rm end}|$, this possibility can be achieved if one has a long period of inflation, with number of $e$-folds:
%\be 
%N_{\rm infl} \geq \(\frac{2 \pi M_{\rm Pl}}{H_{\rm infl}}\)^2 .
%\ee
% 
%\DT{Although it does generate a spatial landscape of gigantic regions with different CC...} 

\begin{figure}[t]
$$\includegraphics[width=1.1\textwidth]{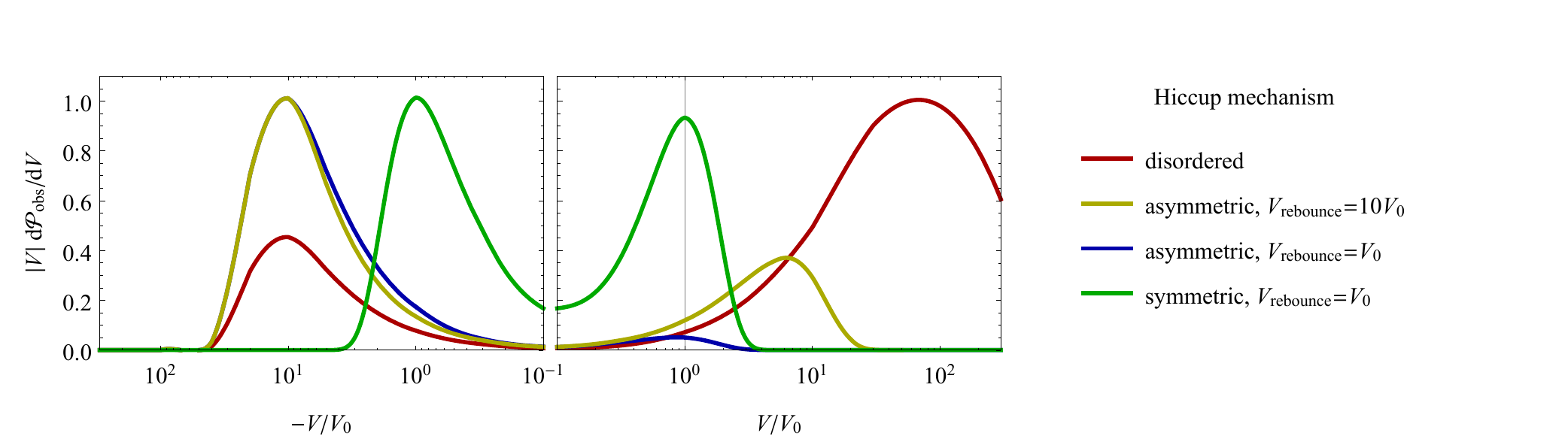}$$
\caption{\em 
\label{fig:CC} Possible probability distribution of eq.\eq{prob}
of the vacuum energy $V$
in units of the observed value $V_0$.
The red curve shows the case of the usual spatial multiverse~\cite{astro-ph/0005265},
that in our context can arise from a disordered hiccupping.
The yellow and blue curves assume
an ordered landscape with asymmetric hiccupping as in eq.\eq{asymmetric}
for different values of $V_{\rm rebounce}$, whereas the green curve assumes a symmetric hiccup.
}
\end{figure}

\medskip

This hiccup mechanism preserves, on average, the value of $V$.
Since the field $\phi$ classically rolls down a bit whenever a cycle
starts with $V > V_{\rm end}$,
$V$ gradually decreases and 
after a large number of cycles the probability distribution
of $V$ becomes, for $V_{\rm rebounce}\gg |V_{\rm end}|$
%\emph{inertia} in the hiccupping mechanism is possible and the a priori probability distribution of the cosmological constant $\mathcal{P}(V)$ can be non-uniform on the relevant scale, in particular peaking at 0.
%\new{At the $i$-th cycle, the probability to have a vacuum energy $V^{(i)}$ is given by $\mathcal{P}(V^{(i)}-V_{\rm final}^{(i-1)})$ where $V_{\rm final}^{(i-1)}$ is the final value of the vacuum energy at the previous cycle, equal to $V_{\rm end}$ or $\simeq V^{(i-1)}$) for cycles that start with positive or negative $V^{(i-1)}$, respectively. By virtue of the asymmetric dynamics due to the relaxation process, along different cycles the vacuum energy decreases on average.  However, the random walk always generates also universes with positive cosmological constant, as the observed one. Numerically, we find that the distribution of universes after a large number of cycles approaches the stationary form:
\be 
\mathcal{P}(V) \propto \left\{ \begin{array}{ll}
1 \qquad & V < - V_{\rm rebounce} \\
e^{- \sfrac{(V+V_{\rm rebounce})^2}{2 \sigma^2}} \qquad & V\geq - V_{\rm rebounce}   
\end{array}
\right. \label{eq:asymmetric}
\ee
with $\sigma \simeq 1.3 V_{\rm rebounce}$ according to
numerical simulations.
We refer to this as \emph{asymmetric} hiccup.

An alternative speculative possibility is that 
the downward average drift of $V$ is avoided by some
\emph{symmetric} hiccup mechanism that gives a distribution of $V$
peaked around the special point of the dynamics $V = V_{\rm end}$
also for cycles with negative vacuum energy
(e.g.\ thermalisation might cause loss of memory).
In such a case $P(V)=P_1(V)$ may be peaked
around some small scale.

%also for cycles with negative vacuum energy, allowing for a symmetric distribution peaked around the special point of the dynamics $V = V_{\rm end} \simeq 0$. In this way, the cosmological constant does not decrease on average and could give, for instance, the distribution $\mathcal{N}(V) = \mathcal{P}(V-V_{\rm end})$ for this speculative \emph{symmetric} hiccup mechanism.}

\subsection{Probability distribution of the cosmological constant}
Finally, we
discuss the probability distribution of the cosmological constant $\mathcal{P}_{\rm obs}(V)$
measured by random observers taking their anthropic selection into account.
In our case $\mathcal{P}_{\rm obs}(V)$ 
is simply given by the product of $\mathcal{P}(V)$
times an astrophysical 
factor $\mathcal{P}_{\rm ant}(V)$
that estimates how many observers form as function of the vacuum energy $V$.
As the volume of a flat universe is infinite, some 
regularising volume $\mathcal{V}_{\rm reg}$ is needed~\cite{gr-qc/9406010,astro-ph/9908115,astro-ph/0002387}:
\be\label{eq:prob}
\mathcal{P}_{\rm obs}(V) \propto \, \mathcal{P}(V) %\lim_{\mathcal{V}_{\rm reg} \to\infty}
\mathcal{P}_{\rm ant}(V),\qquad 
\mathcal{P}_{\rm ant}(V)\propto
 \int d t  \,\mathcal{V}_{\rm reg} \,  \frac{d^2 n_{\rm obs}}{dt \, d \mathcal{V}}(V) .
\ee
The temporal integral is over the finite lifetime of a single cycle. 
The quantity ${d^2 n_{\rm obs}}/{dt \, d \mathcal{V}}$ is the observer production rate per unit time and comoving volume.\footnote{As the
literature is not univocal, we adopt the following choice.
For positive cosmological constants we take the observer production rate from the numerical simulations in~\cite{SmallLambda}, which qualitatively agree with the semi-analytical approach of~\cite{0810.3044}. 
 For the observer model, we choose the ``stellar-formation-rate plus fixed-delay'' model~\cite{0907.4917}, where the rate of formation of observers is taken as proportional to the formation rate of stars, with a 5 Gyr fixed time delay inspired by the formation of complex-enough life on Earth.  
For negative cosmological constants, we approximate the star formation rate as the zero cosmological constant rate supplemented by a hard cut-off at the crunch time of eq.~\eqref{eq:crunch_time}.
In doing so, we neglect a possible new phase of star formation during contraction
 since we assume a fixed time delay $\approx \unit[5]{Gyr}$ for the formation of observers.}
%Eq.~\eqref{eq:prob} is obtained following the evolution of a given comoving volume along the different cycles: the result does not depend on the chosen volume or measure. 
%Since we are not comparing different volumes we do not encounter the
%ambiguities typically present in spatial multiverse calculations.  
The anthropic factor depends on the prescription adopted
to regularise the number of observers. 
Following Weinberg~\cite{astro-ph/0005265} we consider
the number of observers per unit of mass, which corresponds to
${\cal V}_{\rm reg}=1$ in eq.~\eqref{eq:prob}.
This measure prefers vacuum-energy densities 2 orders of magnitude larger than 
the observed $V_0$.
%In the numerical analysis below we will do the same, to assess how the temporal multiverse changes these results. Notice however that 
This unsatisfactory aspect of the standard spatial multiverse can be limited 
by choosing appropriate regularisation volumes, such as the causal-diamond measure
 (see, for instance, \cite{0805.2173,hep-th/0605263,hep-th/0702115,0808.3770}).

Without needing such choices, 
a temporal multiverse can give a probability distribution of $V$
peaked around its observed value.
This needs an ordered landscape with small $V_{\rm rebounce}$.
Figure~\ref{fig:CC} shows numerical result for $P_{\rm obs}(V)$ 
assuming $V_{\rm end} \ll V_0$:
\begin{itemize}
\item The red curve considers a disordered hiccup,
or a ordered hiccup with $V_{\rm rebounce} \gg10^3 V_0$:
they both give a flat $P(V)$ around $V_0$,
reproducing the usual $\Lambda$CDM anthropic selection~\cite{Weinberg:1987dv,Weinberg:2000yb}: vacuum energy densities 2 orders of magnitude larger than $V_0$ are preferred.

\item The yellow and blue curves
assume an ordered asymmetric hiccup, that cuts
large positive values of $V$, but not negative large values.
 
 \item The green curve assumes an ordered symmetric hiccup,
 that cuts large (positive and negative) values of $V$.
Assuming a small $V_{\rm rebounce} \sim V_0$
gives a $\mathcal{P}_{\rm obs}(V)$ peaked around the observed $V_0$,
while 
the measure-dependent 
anthropic factor $\mathcal{P}_{\rm and}(V)$ becomes irrelevant, being
approximatively constant in such a small $V$ interval.

\end{itemize}

%As for the spatial multiverse, by using measures such as the causal diamond, we would obtain preference for the observed value of the cosmological constant also for the temporal multiverse with asymmetric hiccup. Positive values of the vacuum energy are not penalized as long as $V_{\rm rebounce}$ is large enough.

%$\Lambda_0$ (whereas astrophysical considerations only support values higher by 2 orders of magnitude, as in original Weinberg's estimate~\cite{Weinberg:1987dv}, see also~\cite{SmallLambda} for instance). 

\section{Conclusions}\label{concl}
The authors of~\cite{Graham:2019bfu} proposed
a dynamical mechanism that makes the small vacuum energy density observed in cosmology less fine-tuned from the point of view of particle physics.
This possibility was put forward as an alternative to anthropic selection in a multiverse.
However, given that multiple vacua are anyhow needed by the mechanism
of~\cite{Graham:2019bfu}, and
that a multiverse of many vacua is anyhow
suggested by independent considerations,
we explored how the ingredients proposed in~\cite{Graham:2019bfu}
behave in a multiverse context.

A first ingredient of~\cite{Graham:2019bfu}
is a scalar with a bottom-less potential and small slope that relaxes the cosmological constant down to small negative values.
In section~\ref{phi} we computed the resulting cosmology. 
In particular, we found that any universe  eventually undergoes a phase of contraction, leading to a crunch, even starting from a positive cosmological constant. 
This avoids the possible Boltzmann-brain paradox 
generated e.g.\ by the observed positive cosmological constant.
%that affects many measures~\cite{hep-th/0208013,hep-th/0510003,hep-th/0610132}. 
We calculated the  parameter space compatible with present observations, with the novel behaviour starting in the future. 

A second ingredient of~\cite{Graham:2019bfu} is a  mechanism that rebounces a contracting universe into an expanding one and mildly changes its cosmological constant. In section~\ref{tmul} we explored how rebounces would 
affects attempts of computing  probabilities in the multiverse. In particular
a steady-state temporal multiverse becomes possible,
as anti-de Sitter vacua are no longer terminal and rebounce into
expanding regions.

%The necessity of the contraction phase also for positive values of the cosmological constants strongly affects the dynamics of the multiverse and can change it also qualitatively. In particular, for both positive and negative cosmological constants

In section~\ref{RR} we combined both ingredients above.
Any region now undergoes cycles of expansion, contraction and rebounce
in a finite time\footnote{This turns out to be similar, in spirit, to the proposal of Ref.~\cite{astro-ph/0605173} based on the Ekpyrotic Universe, where periodic brane collisions give rise to cyclic big bangs that reheat the Universe during Abbott's relaxation, thus solving its empty-Universe problem. However, in our case relaxation and reheating are intertwined: the latter is triggered by the contraction phase due to the former.}.
This temporal universe is not affected by issues that often plague the spatial multiverse.
For instance, the Boltzmann-brain paradox and the youngness paradox~\cite{hep-th/0702178} are avoided because there are no exponentially inflating regions nucleating habitable universes.
In a part of its parameter space, the mechanism of \cite{Graham:2019bfu}
can provide the most likely source of universes 
with vacuum energy density below anthropic boundaries.
One can devise specific models where the probability distribution
of the vacuum energy improves on the situation present in the usual 
anthropic selection,
where the most likely value of the cosmological constant seems
2 orders of magnitude above its observed value.

\footnotesize

\subsubsection*{Acknowledgements}
This work was supported by the ERC grant NEO-NAT.
We thank Dario Buttazzo,  Juan Garcia-Bellido, Luca Di Luzio, Antonio Riotto and  Enrico Trincherini for discussions.


\begin{thebibliography}{nnn}\bibitem{1403.4226}
\article[1403.4226]{A. Salvio, A. Strumia}{JHEP}{1406}{080}{2014}
{Agravity}.


\bibitem{Weinberg:1987dv}
\article[Weinberg:1987dv]{S. Weinberg}{Phys. Rev. Lett.}{59}{2607}{1987}
{Anthropic Bound on the Cosmological Constant}.


\bibitem{astro-ph/9908115}
\article[astro-ph/9908115]{J. Garriga, A. Vilenkin}{Phys. Rev.}{D61}{083502}{1999}
{On likely values of the cosmological constant}.


\bibitem{astro-ph/0005265}
\heparticle[astro-ph/0005265]{S. Weinberg}{The Cosmological constant problems}.


\bibitem{0705.0898}
\article[0705.0898]{J.A. Peacock}{Mon. Not. Roy. Astron. Soc.}{379}{1067}{2007}
{Testing anthropic predictions for Lambda and the CMB temperature}.


\bibitem{0805.2173}
\article[0805.2173]{A. De Simone, A.H. Guth, M.P. Salem, A. Vilenkin}{Phys. Rev.}{D78}{063520}{2008}
{Predicting the cosmological constant with the scale-factor cutoff measure}.


\bibitem{hep-th/0605263}
\article[hep-th/0605263]{R. Bousso}{Phys. Rev. Lett.}{97}{191302}{2006}
{Holographic probabilities in eternal inflation}.


\bibitem{hep-th/0702115}
\article[hep-th/0702115]{R. Bousso, R. Harnik, G.D. Kribs, G. Perez}{Phys. Rev.}{D76}{043513}{2007}
{Predicting the Cosmological Constant from the Causal Entropic Principle}.


\bibitem{0808.3770}
\article[0808.3770]{R. Bousso, B. Freivogel, I-S. Yang}{Phys. Rev.}{D79}{063513}{2008}
{Properties of the scale factor measure}.


\bibitem{Graham:2019bfu}
\heparticle[Graham:2019bfu]{P.W. Graham, D.E. Kaplan, S. Rajendran}{Relaxation of the Cosmological Constant}.


\bibitem{1608.05715}
\article[1608.05715]{L. Alberte, P. Creminelli, A. Khmelnitsky, D. Pirtskhalava, E. Trincherini}{JHEP}{1612}{022}{2016}
{Relaxing the Cosmological Constant: a Proof of Concept}.


\bibitem{mq}
\article[hep-ph/9707380]{V. Agrawal, S.M. Barr, J.F. Donoghue, D. Seckel}{Phys. Rev.}{D57}{5480}{1997}
{The Anthropic principle and the mass scale of the standard model}.
\article[0712.2454]{L.J. Hall, Y. Nomura}{Phys. Rev.}{D78}{035001}{2007}
{Evidence for the Multiverse in the Standard Model and Beyond}.
\article[hep-ph/0703219]{S.M. Barr, A. Khan}{Phys. Rev.}{D76}{045002}{2007}
{Anthropic tuning of the weak scale and of $m_u / m_d$ in two-Higgs-doublet models}.
\article[0809.1647]{R.L. Jaffe, A. Jenkins, I. Kimchi}{Phys. Rev.}{D79}{065014}{2008}
{Quark Masses: An Environmental Impact Statement}.


\bibitem{1906.00986}
\heparticle[1906.00986]{G. D'Amico, A. Strumia, A. Urbano, W. Xue}{Direct anthropic bound on the weak scale from supernovae explosions}.


\bibitem{astro-ph/9401042}
\article[astro-ph/9401042]{J. Garcia-Bellido}{Nucl. Phys.}{B423}{221}{1994}
{Jordan-Brans-Dicke stochastic inflation}.


\bibitem{1807.06209}
\heparticle[1807.06209]{{\sc Planck} Collaboration}{Planck 2018 results. VI. Cosmological parameters}.


\bibitem{PDG} \article[Tanabashi:2018oca]{{\sc PDG} Collaboration}{Phys. Rev.}{D98}{030001}{2018}
{Review of Particle Physics}.


\bibitem{hep-th/0208013}
\article[hep-th/0208013]{L. Dyson, M. Kleban, L. Susskind}{JHEP}{0210}{011}{2002}
{Disturbing implications of a cosmological constant}.


\bibitem{hep-th/0510003}
\article[hep-th/0510003]{D.N. Page}{J. Korean Phys. Soc.}{49}{711}{2005}
{The Lifetime of the universe}.


\bibitem{hep-th/0610132}
\article[hep-th/0610132]{R. Bousso, B. Freivogel}{JHEP}{0706}{018}{2006}
{A Paradox in the global description of the multiverse}.


\bibitem{hep-th/0206228}
\article[hep-th/0206228]{G.T. Horowitz, J. Polchinski}{Phys. Rev.}{D66}{103512}{2002}
{Instability of space-like and null orbifold singularities}.


\bibitem{Garriga:2005av}
\article[hep-th/0509184]{J. Garriga, D. Schwartz-Perlov, A. Vilenkin, S. Winitzki}{JCAP}{0601}{017}{2005}
{Probabilities in the inflationary multiverse}.


\bibitem{hep-th/0702178}
\article[hep-th/0702178]{A.H. Guth}{J. Phys.}{A40}{6811}{2007}
{Eternal inflation and its implications}.


\bibitem{0808.3778}
\article[0808.3778]{A. De Simone, A.H. Guth, A.D. Linde, M. Noorbala, M.P. Salem, A. Vilenkin}{Phys. Rev.}{D82}{063520}{2008}
{Boltzmann brains and the scale-factor cutoff measure of the multiverse}.


\bibitem{0812.0005}
\article[0812.0005]{A.D. Linde, V. Vanchurin, S. Winitzki}{JCAP}{0901}{031}{2008}
{Stationary Measure in the Multiverse}.


\bibitem{0901.4806}
\article[0901.4806]{R. Bousso}{Phys. Rev.}{D79}{123524}{2009}
{Complementarity in the Multiverse}.


\bibitem{1005.2783}
\article[1005.2783]{R. Bousso, B. Freivogel, S. Leichenauer, V. Rosenhaus}{Phys. Rev.}{D82}{125032}{2010}
{Boundary definition of a multiverse measure}.


\bibitem{1104.2324}
\article[1104.2324]{Y. Nomura}{JHEP}{1111}{063}{2011}
{Physical Theories, Eternal Inflation, and Quantum Universe}.


\bibitem{0901.2644}
\article[0901.2644]{Y-S. Piao}{Phys. Lett.}{B677}{1}{2009}
{Proliferation in Cycle}.


\bibitem{1210.7540}
\article[1210.7540]{J. Garriga, A. Vilenkin}{JCAP}{1305}{037}{2013}
{Watchers of the multiverse}.


\bibitem{Graham:2017hfr}
\article[1709.01999]{P.W. Graham, D.E. Kaplan, S. Rajendran}{Phys. Rev.}{D97}{044003}{2018}
{Born again universe}.


\bibitem{Abbott:1984qf}
\article[Abbott:1984qf]{L.F. Abbott}{Phys. Lett.}{150B}{427}{1985}
{A Mechanism for Reducing the Value of the Cosmological Constant}.


\bibitem{1504.07551}
\article[1504.07551]{P.W. Graham, D.E. Kaplan, S. Rajendran}{Phys. Rev. Lett.}{115}{221801}{2015}
{Cosmological Relaxation of the Electroweak Scale}.


\bibitem{1609.06320}
\article[1609.06320]{A. Arvanitaki, S. Dimopoulos, V. Gorbenko, J. Huang, K. Van Tilburg}{JHEP}{1705}{071}{2017}
{A small weak scale from a small cosmological constant}.


\bibitem{1801.03926}
\article[1801.03926]{J.M. Cline, J.R. Espinosa}{Phys. Rev.}{D97}{035025}{2018}
{Axionic landscape for Higgs coupling near-criticality}.


\bibitem{1809.07338}
\heparticle[1809.07338]{M. Geller, Y. Hochberg, E. Kuflik}{Inflating to the Weak Scale}.


\bibitem{1811.12390}
\heparticle[1811.12390]{C. Cheung, P. Saraswat}{Mass Hierarchy and Vacuum Energy}.


\bibitem{1904.00020}
\heparticle[1904.00020]{A. Hook, J. Huang}{Searches for other vacua I: bubbles in our universe}.


\bibitem{1511.00132}
\article[1511.00132]{K. Choi, S.H. Im}{JHEP}{1601}{149}{2016}
{Realizing the relaxion from multiple axions and its UV completion with high scale supersymmetry}.


\bibitem{1511.01827}
\article[1511.01827]{D.E. Kaplan, R. Rattazzi}{Phys. Rev.}{D93}{085007}{2016}
{Large field excursions and approximate discrete symmetries from a clockwork axion}.


\bibitem{1610.07962}
\article[1610.07962]{G.F. Giudice, M. McCullough}{JHEP}{1702}{036}{2017}
{A Clockwork Theory}.


\bibitem{1802.01591}
\article[1802.01591]{D. Teresi}{Phys. Lett.}{B783}{1}{2018}
{Clockwork without supersymmetry}.


\bibitem{1703.02576}
\article[1703.02576]{K. Saikawa}{Universe}{3}{40}{2017}
{A review of gravitational waves from cosmic domain walls}.


\bibitem{1205.6260}
\article[1205.6260]{M. Pospelov, S. Pustelny, M.P. Ledbetter, D.F. Jackson Kimball, W. Gawlik, D. Budker}{Phys. Rev. Lett.}{110}{021803}{2013}
{Detecting Domain Walls of Axionlike Models Using Terrestrial Experiments}.


\bibitem{gr-qc/9406010}
\article[gr-qc/9406010]{A. Vilenkin}{Phys. Rev. Lett.}{74}{846}{1994}
{Predictions from quantum cosmology}.


\bibitem{astro-ph/0002387}
\article[astro-ph/0002387]{S. Weinberg}{Phys. Rev.}{D61}{103505}{2000}
{A Priori probability distribution of the cosmological constant}.


\bibitem{SmallLambda}
See e.g.\
\article[1801.08781]{L.A. Barnes, P.J. Elahi, J. Salcido, R.G. Bower, G.F. Lewis, T. Theuns, M. Schaller, R.A. Crain, J. Schaye}{Mon. Not. Roy. Astron. Soc.}{477}{3727}{2018}
{Galaxy Formation Efficiency and the Multiverse Explanation of the Cosmological Constant with EAGLE Simulations}.


\bibitem{0810.3044}
\article[0810.3044]{R. Bousso, S. Leichenauer}{Phys. Rev.}{D79}{063506}{2008}
{Star Formation in the Multiverse}.


\bibitem{0907.4917}
\article[0907.4917]{R. Bousso, S. Leichenauer}{Phys. Rev.}{D81}{063524}{2009}
{Predictions from Star Formation in the Multiverse}.


\bibitem{Weinberg:2000yb}
\heparticle[Weinberg:2000yb]{S. Weinberg}{The Cosmological constant problems}.


\bibitem{astro-ph/0605173}
\article[astro-ph/0605173]{P.J. Steinhardt, N. Turok}{Science}{312}{1180}{2006}
{Why the cosmological constant is small and positive}.


\end{thebibliography}
\end{document}